\documentclass[10pt,journal,compsoc]{IEEEtran}
\usepackage{amsmath,amssymb,amsfonts}
\usepackage{algorithmic}
\usepackage{graphicx}
\usepackage{textcomp}
\usepackage{xcolor}
\usepackage{algorithm}

\usepackage{booktabs} 

\usepackage[autolinebreaks,useliterate]{mcode}
\usepackage{hyperref}

\ifCLASSOPTIONcompsoc
  \usepackage[nocompress]{cite}
\else
  \usepackage{cite}
\fi
\ifCLASSINFOpdf
\else
\fi
\begin{document}
%
\title{STD: A Seasonal-Trend-Dispersion Decomposition of Time Series}
%
%
%
%

\author{Grzegorz Dudek
\IEEEcompsocitemizethanks{\IEEEcompsocthanksitem 
G. Dudek was with the Department of Electrical Engineering, Czestochowa University of Technology, 42-200 Czestochowa, Al. Armii Krajowej 17, Poland.\protect\\
E-mail: grzegorz.dudek@pcz.pl}
\thanks{Manuscript received ..., 2021; revised ..., 2022.}}

%
%

\markboth{Journal of \LaTeX\ Class Files,~Vol.~14, No.~8, August~2015}%
{Shell \MakeLowercase{\textit{et al.}}: Bare Demo of IEEEtran.cls for Computer Society Journals}
%



\IEEEtitleabstractindextext{%
\begin{abstract}
The decomposition of a time series is an essential task that helps to understand its very nature. It facilitates the analysis and forecasting of complex time series expressing various hidden components such as the trend, seasonal components, cyclic components and irregular fluctuations. Therefore, it is crucial in many fields for forecasting and decision processes. In recent years, many methods of time series decomposition have been developed, which extract and reveal different time series properties. Unfortunately, they neglect a very important property, i.e. time series variance. To deal with heteroscedasticity in time series, the method proposed in this work -- a seasonal-trend-dispersion decomposition (STD) -- extracts the trend, seasonal component and component related to the dispersion of the time series. We define STD decomposition in two ways: with and without an irregular component. We show how STD can be used for time series analysis and forecasting.
\end{abstract}

}


\maketitle

\IEEEdisplaynontitleabstractindextext

%
\IEEEpeerreviewmaketitle

\IEEEraisesectionheading{\section{Introduction}\label{Intro}}

%
%
%
%
\IEEEPARstart{A}{time} series expresses states of a certain variable that describe a given phenomenon (economic, biological, physical, etc.) 
observed in subsequent periods. Time series analysis and forecasting is an extremely important task in many fields, including business, industry, government, politics, health and medicine \cite{Pal05}. However, this task can be difficult due to the complex nature of the time series. Time series can exhibit a variety of unobservable (latent) components that can be associated with
different types of temporal variations. These include:
(1) a long-term tendency or trend, (2) cyclical movements superimposed upon the long-term trend (usually non-periodical), (3) seasonal variations (periodical), and (4) irregular fluctuations. In economics, 
the seasonal variations represent the composite effect of climatic and institutional events which repeat more or less regularly each year \cite{Dag16}. The cycles appear to reach their peaks during periods of economic prosperity and their troughs during periods of depression. Their rise and fall constitute the business cycle.   

Extracting the components of a time series can help us to understand the underlying process and to forecast it. Instead of building a complex forecasting model for the composed time series, after decomposition into basic components, we can built simpler specialized models for each component. This approach is very common in forecasting using both classical statistical methods and machine learning methods. Therefore, many methods of time series decomposition have been proposed.        

\subsection{Related Work}

Time series decomposition has a long history dating back to the mid 19th century \cite{Maz18}. The idea of decomposing the time series into unobservable components appeared in the work of 19th century economists who drew their inspiration from astronomy and meteorology \cite{Buy47}. Much research back then was done to reveal the "cycles" that made it possible to explain and predict economic crises. In 1884, Poynting proposed price averaging as a tool to eliminate trend and seasonal fluctuations \cite{Mak98}. Later his approach was extended by other researchers including Copeland who was the first to attempt to extract the seasonal component \cite{Dok20}. Persons was the first to define the various components of a time series, i.e. the trend, cycle, seasonal and irregular components, and proposed an algorithm to estimate them (link relatives method) \cite{Per19}. The process of decomposition was refined by Macauley who proposed a way of smoothing time series, which has become a classic over time \cite{Mac31}. Based on Macauley's method, the Census II method was developed and its numerous variants are widely used today such as X-11, X-11-ARIMA, X-12-ARIMA, X-13ARIMA-SEATS, and TRAMO-SEATS. A detailed discussion of these methods is provided by \cite{Dag16}.

Structural time series decomposition, which involves decomposing a series into components having a direct interpretation, is very useful from a practical point of view. A structural model is formulated directly in terms of unobserved components, such as the trend, cycles, seasonals and remaining component. These components can be combined additively or multiplicatively. An additive decomposition is applied if the variation around the trend-cycle, or the magnitude of seasonal variations, does not change with the time series level. When such variation is observed to be proportional to the time series level, multiplicative decomposition is more appropriate.

To extract the components of the series, both parametric or non-parametric methods are used. A parametric approach imposes a specific model on the component, e.g. linear or polynomial. The nonparametric approach offers more possibilities because it does not limit the model to a specific class. A popular example of a non-parametric method to extract a trend is smoothing with a moving average.

One of the most widely used methods of time series decomposition is STL (Seasonal and Trend decomposition using Loess) \cite{Cle90}. STL is additive. The STL decomposition procedure is iterative and relies on the alternate estimation of the trend and the seasonal components using locally estimated scatterplot smoothing (Loess), which can estimate nonlinear relationships. The seasonal component is allowed to change over time. It is composed of seasonal patterns estimated based on $k$ consecutive seasonal cycles, where $k$ controls how rapidly the seasonal component can change. Other attractive features of STL are: robustness to outliers and missing data, the ability to decompose time series with seasonality of any frequency, and the possibility of implementation using numerical methods instead of mathematical modeling.

Another popular method of additive time series decomposition uses a discrete wavelet transform. Wavelet-based multi-resolution analyses decomposes the series in an iterative process into components with decreasing frequencies \cite{Mal89}.
In the subsequent levels of decomposition, the series is processed by a pair of filters -- high-pass and low-pass (two-channel subband coding). The result is a low-frequency component, the so-called approximation, representing the trend and a high-frequency component, the so-called detail, representing the detailed features of the series. In each iteration, the approximation from the previous iteration is decomposed into detail and new approximation. The sum of all the details produced at all levels, and the lowest-level approximation gives the input series. The decomposition depends on the form and parameters of the wavelet function, which is a function of both time and frequency. 


In \cite{Hua98}, Empirical Mode Decomposition (EMD) was proposed, which decomposes the time series in the time domain into components called Intrinsic Mode Functions (IMFs). These form a complete and nearly orthogonal basis for the original time series. An IMF amplitude and frequency can vary with time.
The IMFs are obtained by applying a recursive so-called sifting process. This extracts the local minima and maxima of the series and then interpolates them separately using cubic splines.
The IMFs extracted at subsequent levels are characterized by ever lower frequencies. 
Since the decomposition is based on the local characteristic time scale of the data, EMD is suitable for both non-linear and non-stationary time series.

Other, less popular, time series decomposition methods include: Variational Mode Decomposition (VMD) \cite{Dra14}, Singular Spectrum Analysis (SSA) \cite{Gol01}, and Seasonal-Trend Decomposition based on Regression (STR) \cite{Dok20}. VMD is a generalization of the classical Wiener filter into many adaptive signal bands. It extracts a set of IMFs defined in different frequency bands, which optimally reconstruct the time series. As an alternative to EMD, VMD is devoid of some EMD limitations, such as the lack of theoretical foundations, sensitivity to sampling and data disturbance, and the dependence of the result on the methods of extremes detection and  envelope interpolation. 

SSA is based on the matrix representation of the time series in the form of a so-called trajectory matrix (Hankel matrix) and its singular value decomposition (SVD). Using the SVD products, i.e. eigentriples, the trajectory matrix is expressed as the sum of elementary matrices. The time series components are obtained by appropriate grouping of the elementary matrices using eigentriples for this purpose. The SSA decomposition is additive. The components obtained as a result are interpretable. They express the trend, periodic components and random disturbances.

STR is an additive decomposition with a matrix representation of the seasonal component. The method can produce multiple seasonal and cyclic components. Seasonal components 
can be fractional, flexible over time, and can have complex topology. STR allows us to take into account the influence of additional external variables on decomposition and to estimate confidence intervals for components.




\subsection{Motivation and Contribution} 

Existing methods of time series decomposition extract different components expressing different time series properties. 
However, to our knowledge, none of them extracts the component representing the series dispersion. 
To fill this gap, this work proposes a new method of time series decomposition that extracts the components of the trend, seasonality and dispersion. It can be useful for analysis and forecasting of heteroscedastic time series. 

Our research contributions can be summarized as follows:

\begin{enumerate}
\item We propose a new method of time series decomposition. It has two variants. In the first, STD, it extracts the trend, seasonal component and dispersion component. In the second variant, STDR, it extracts additionally an irregular component (reminder).
\item We demonstrate how the proposed decomposition method can be used for simplifying and solving complex forecasting problems including those with multiple seasonality and variable variance. 
\end{enumerate}

The rest of the work is organized as follows. Section 2 describes decomposition of the heteroscedastic time series using standard methods. The proposed STD and STDR methods are presented in Section 3. Section 4 gives some application examples and shows how STD can be used for forecasting. Finally, Section 5 concludes the work.


\section{Decomposition of Heteroscedastic Time Series using Additive and Multiplicative Methods}

Typically, time series decomposition can be expressed in an additive or multiplicative form as follows \cite{Hyn22,Dag16}:
\begin{equation}
y_t = T_t + S_t + R_t
\label{eqd0}
\end{equation}
\begin{equation}
y_t = T_t \times S_t \times R_t
\label{eqd01}
\end{equation}
where $y_t$ denotes the observed series, $T_t$ is a trend-cycle component combining the trend and cycle (often just called the trend for simplicity), $S_t$ is the seasonal component, and 
$R_t$ is the irregular component (reminder), all at period $t$.
 
In the additive model, heteroscedasticity in $y_t$ has to be expressed by heteroskadisticity in one or more decomposition products. Usually, the trend is a smoothed original time series, so it does not include short-term variations of varying variance. These variations appear in the seasonal and/or irregular components. If the decomposition method produces a regular seasonal component, i.e. composed of the seasonal cycles of the same shape, which is a classical approach \cite{Hyn22}, the time series variance has to be expressed by the irregular component. But a desired property of the irregular component, which is often assumed for inferential purposes, is to be normally identically distributed and not correlated, which implies independence \cite{Dag16}. Hence, $R_t \sim NID(0,\sigma^2)$. When the variance of the irregular component changes in time, it does not express a white noise in the strict sense. Therefore, the additive model \eqref{eqd0} is not recommended for heteroscedastic time series.

In the multiplicative model, all components are multiplied, so the variations included in the irregular and seasonal components are amplified or weakened by the trend. An increasing trend increases these variations, while a decreasing trend decreases them. Thus, the multiplicative model is most useful when the variation in time series is proportional to the level of the series.

\begin{figure}[h]
	\centering
	\includegraphics[width=0.280\textwidth]{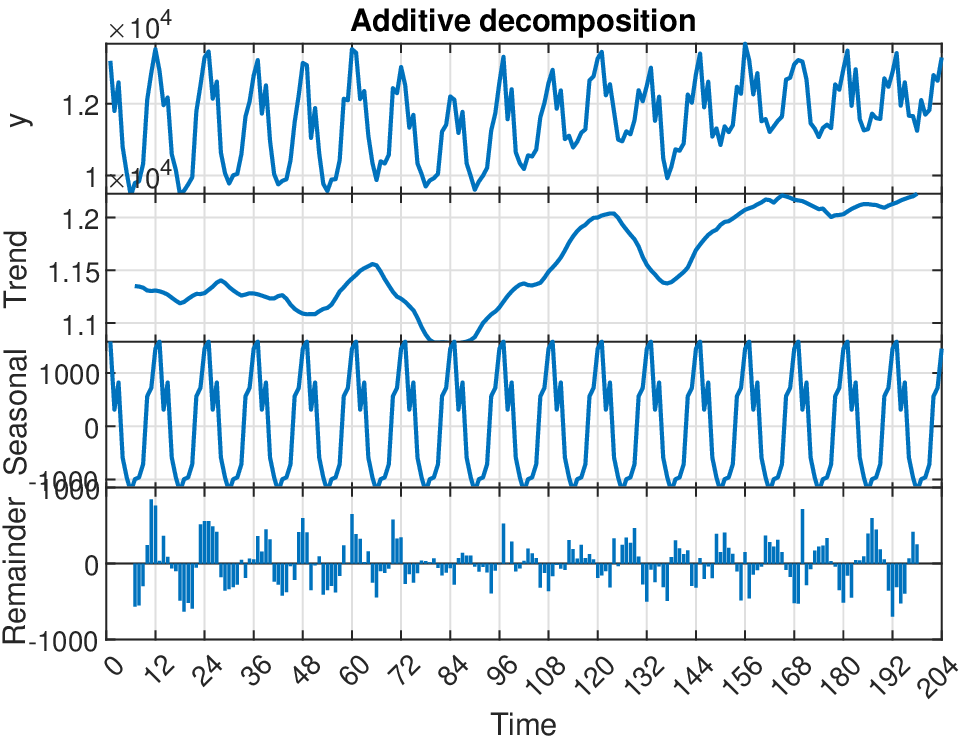}
	\includegraphics[width=0.1432\textwidth]{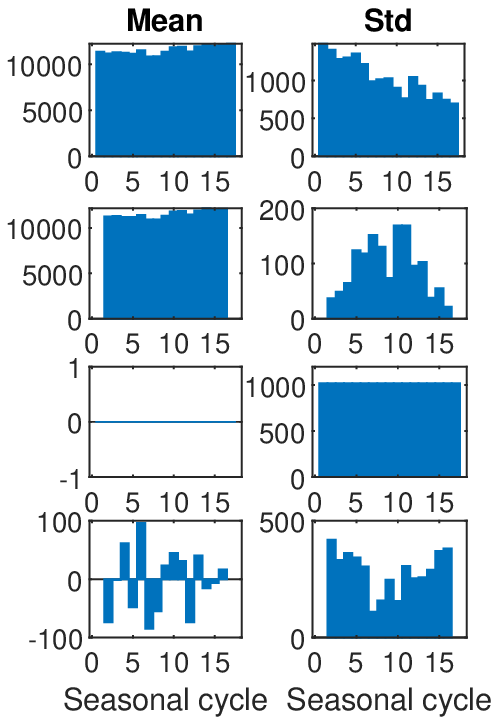}
	\includegraphics[width=0.280\textwidth]{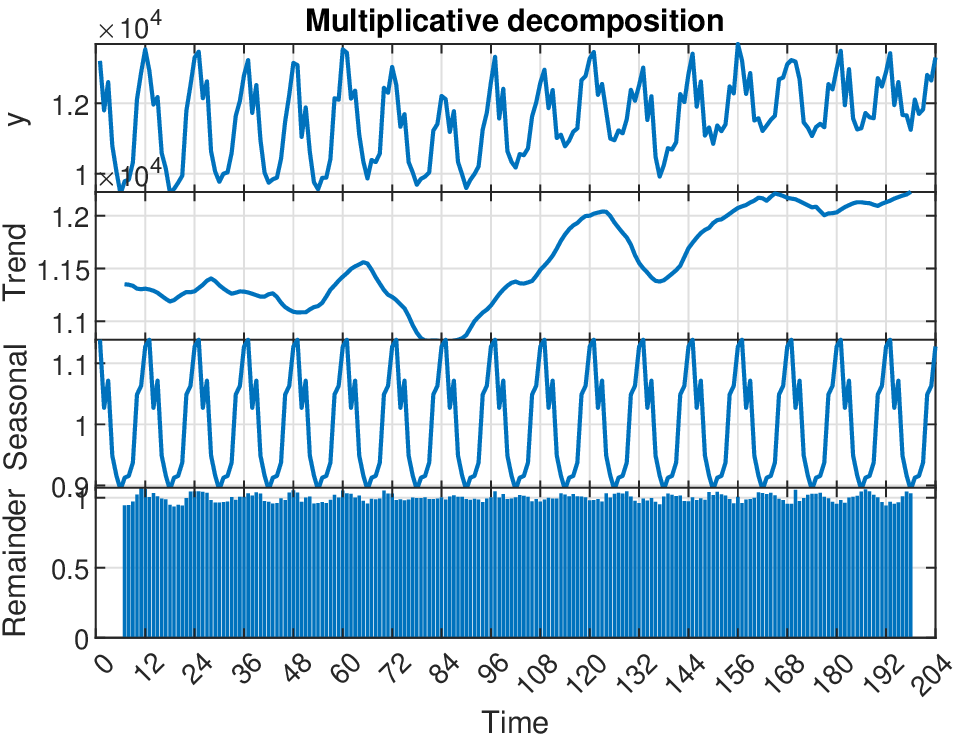}
	\includegraphics[width=0.1432\textwidth]{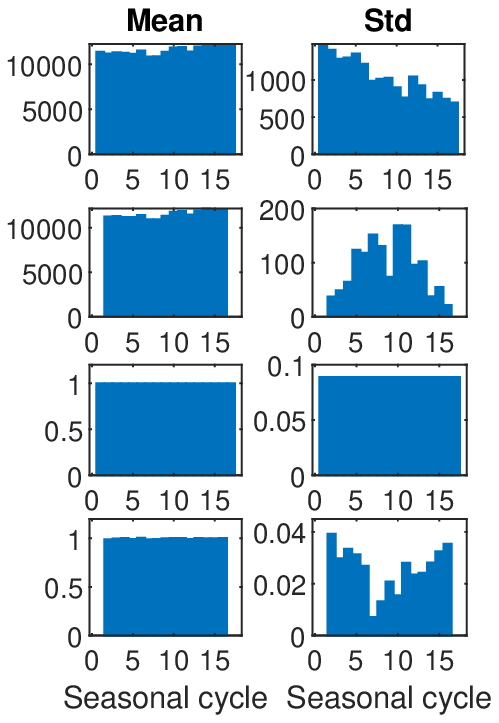}
	\includegraphics[width=0.280\textwidth]{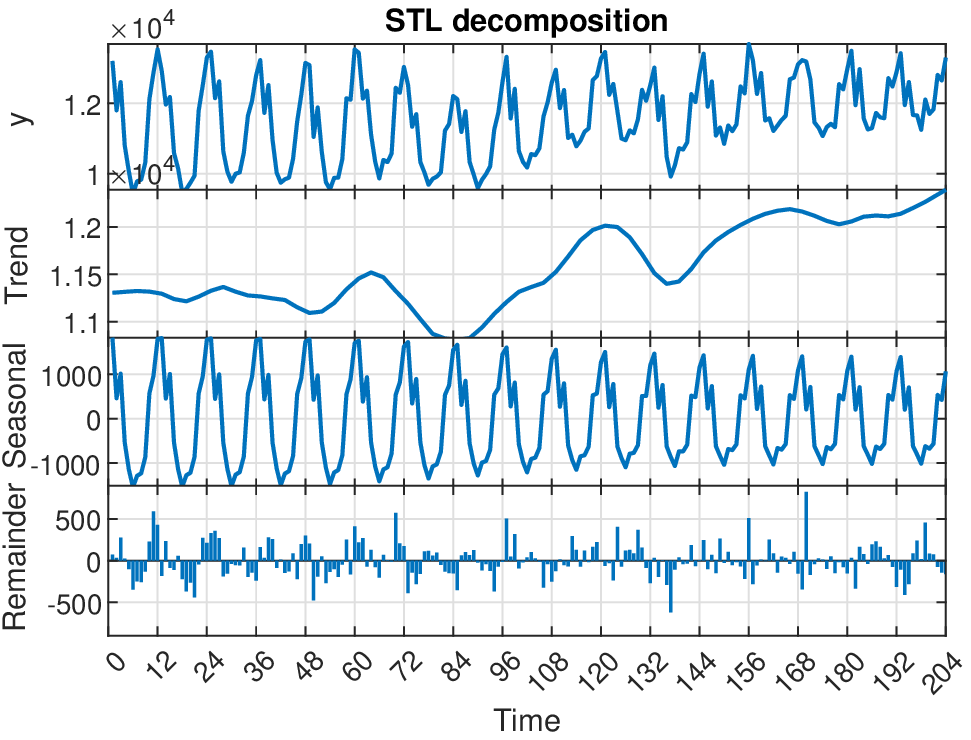}
	\includegraphics[width=0.1432\textwidth]{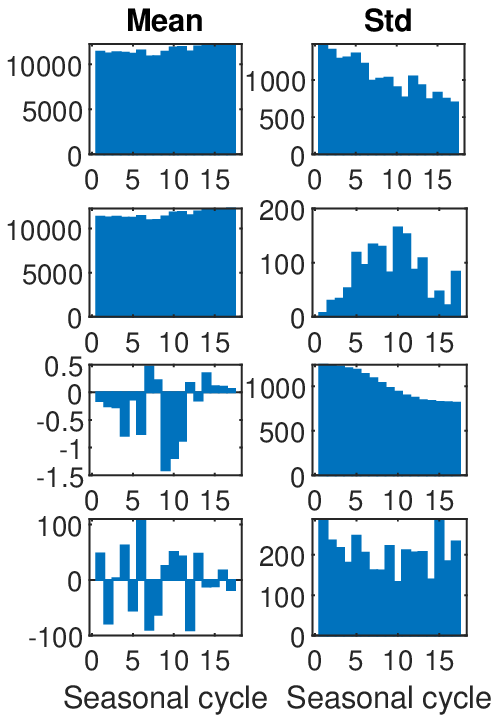}	
	\includegraphics[width=0.280\textwidth]{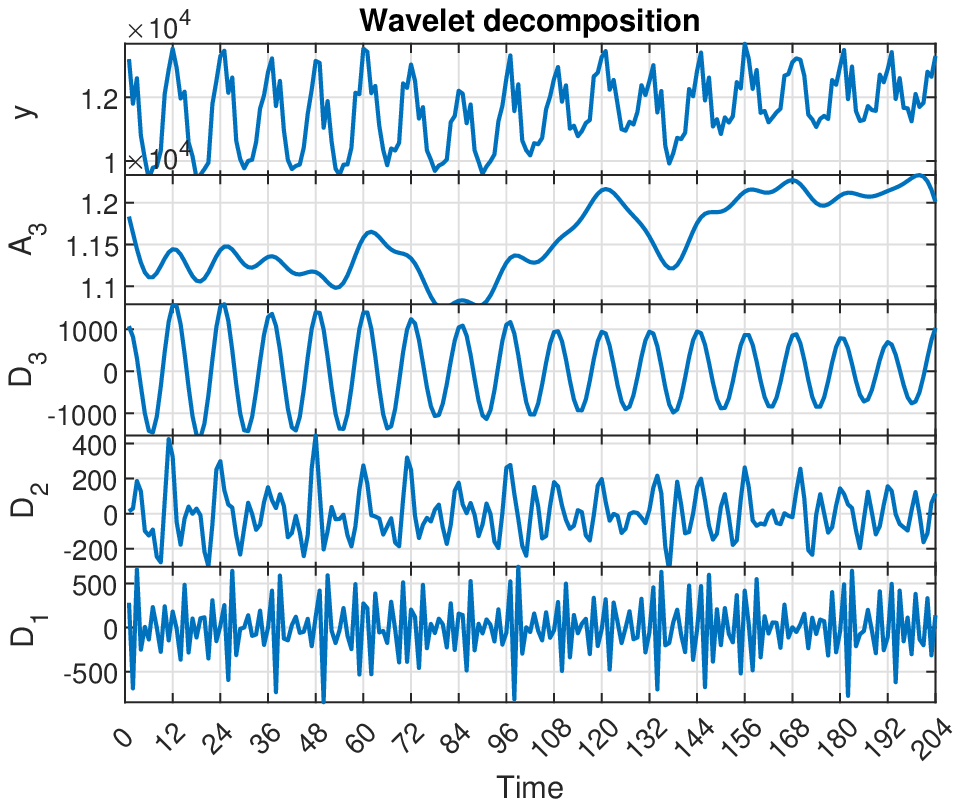}
	\includegraphics[width=0.1432\textwidth]{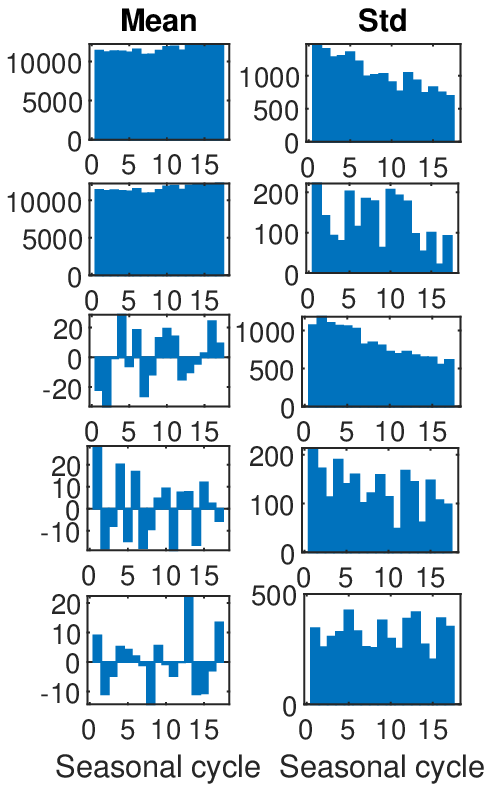}	
	\includegraphics[width=0.280\textwidth]{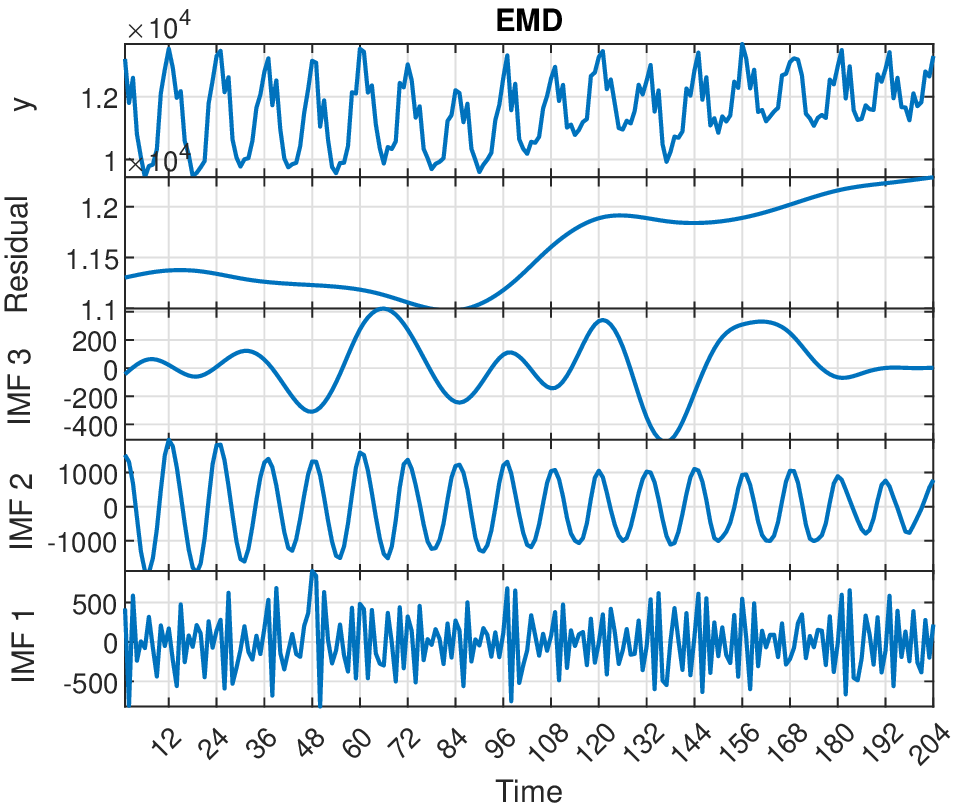}
	\includegraphics[width=0.1432\textwidth]{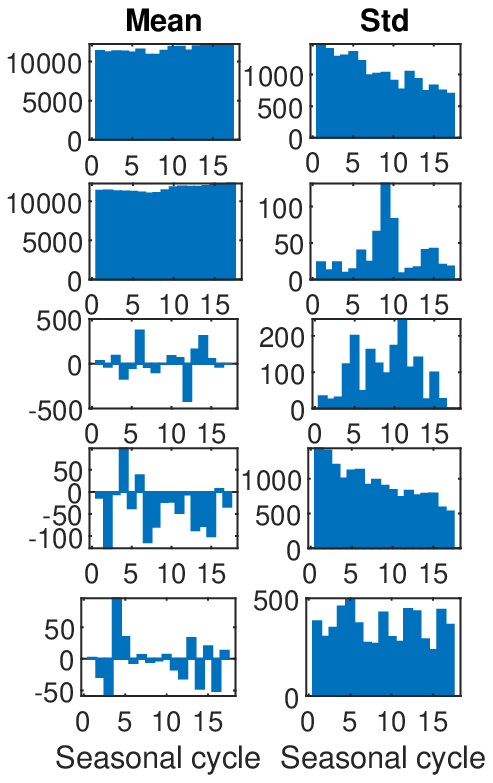}	
	\caption{Monthly electricity demand time series decomposition using standard methods.} 
	\label{Dc}
\end{figure}

Fig. \ref{Dc} shows decomposition of a time series expressing monthly electricity demand for Poland (17 years, observed from 1997 to 2014) using the most popular decomposition methods, i.e. classical additive and multiplicative methods, STL, wavelet transform, and EMD. Note that the times series has decreasing variations with the trend level. Mean values of the series and their standard deviations are shown in the bar chars shown in the right panel. They are calculated for successive sequences of length $n=12$. 
To estimate the trend, the classical additive and multiplicative methods use two-sided moving averages. The negative effect of this is that the first and last few observations are missing from the trend and irregular components. The classical methods assume that the seasonal component is constant throughout the entire series. This constant seasonal pattern is determined as an average of all seasonal sequences of the detrended series. The long-term variability is expressed by the trend. Note how this variability changes over time in the std-chart. The short-term variability is expressed in the remainder component. The std-chart for this component shows that the variance is smallest in the middle part of the data period. In this part, the combined trend and seasonal components approximate the time series most accurately. In the first part, the amplitude of the combined components is smaller than the amplitude of the real series and must be increased by the irregular component. In this part, the extremes of the irregular component correspond to the extremes of the seasonal component. In the final part of the series, the amplitude of the combined trend-seasonal  component is higher that the real amplitude. The irregular component compensates the amplitude. Its extremes are opposite to the extremes of the seasonal component. The compensation function of the irregular component results in its variable variance and autocorrelation.

STL produces a smoother trend than classical decomposition methods due to the use of local polynomial regression. A seasonal component in STL averages the real seasonal patterns but can still reflects its amplitude. Therefore, to compensate for the amplitude mismatch, the irregular component may be smaller than in classical decomposition methods. However, it still expresses the variable variance and autocorrelation. 

Wavelet decomposition produces the components corresponding to the trend ($A_3$) and smoothed seasonal variations ($D_3$) as well as components expressing more detailed variations. Each of them expresses changing variance. As can be seen from Fig. \ref{Dc}, EMD produces the most smoothed trend (residual component) compared to other methods and a separate component representing non-periodical cyclical movements (IMF3). The seasonal component, IMF2, which is very similar to the $D_3$ component generated by wavelet transform, smooths the seasonal cycles significantly. The random component, IMF1, is very similar to the highest-level detail of the wavelet decomposition, $D_1$. The variance of the series is distributed between EMD components.    

Note that the time series variance is not expressed explicitly in the decomposition products of the presented methods. It is hidden in the components. A separate dispersion component could be very useful for time series analysis and forecasting. In the next section, we propose a method which extracts this component.




\section{Seasonal-Trend-Dispersion Decomposition}

Let $\{y_t\}_{t=1}^N$ be a time series with a seasonality of period $n$.
Assume that the length of the series is a multiple of the seasonal period, i.e. $N/n=K, K \in \mathbb{N}$. Time series $y_t$ can be  written as a series of successive seasonal sequences:

\begin{equation}
\{\{y_{i,j}\}_{j=1}^n\}_{i=1}^{K}=\{\{y_{1,j}\}_{j=1}^n, ...,\{y_{K,j}\}_{j=1}^n\} 
\label{yij}
\end{equation}
where $i = 1, ..., K$ is the running number of the seasonal cycle, and $j = 1, ..., n$ is the 
time index inside the given seasonal cycle. The global time index $t = n (i-1) + j$.

The average value of the $i$-th seasonal sequence is:

\begin{equation}
\bar{y}_i=\frac{1}{n}\sum_{j=1}^n y_{i,j}
\label{my}
\end{equation}
and its diversity measure is defined as:

\begin{equation}
\tilde{y}_i=\sqrt{\sum_{j=1}^n (y_{i,j}-\bar{y}_i)^2}
\label{dy}
\end{equation}

The trend component is defined using averages of the seasonal sequences as follows:

\begin{equation}
\{T_t\}_{t=1}^N=\{\{\underbrace{\bar{y}_i, ..., \bar{y}_i}_\text{$n$ times}\}\}_{i=1}^{K}
\label{Tr}
\end{equation}
while the dispersion component is defined using diversities of these sequences:

\begin{equation}
\{D_t\}_{t=1}^N=\{\{\underbrace{\tilde{y}_i, ..., \tilde{y}_i}_\text{$n$ times}\}\}_{i=1}^{K}
\label{Di}
\end{equation}

Based on the trend and dispersion components, we define the seasonal component:

\begin{equation}
S_t=\frac{y_t-T_t}{D_t}
\label{Se}
\end{equation}

The proposed STD decomposition is expressed as follows:

\begin{equation}
y_t = S_t \times D_t+T_t
\label{eqd1}
\end{equation}

\begin{figure}[ht]
	\centering
	\includegraphics[width=0.321\textwidth]{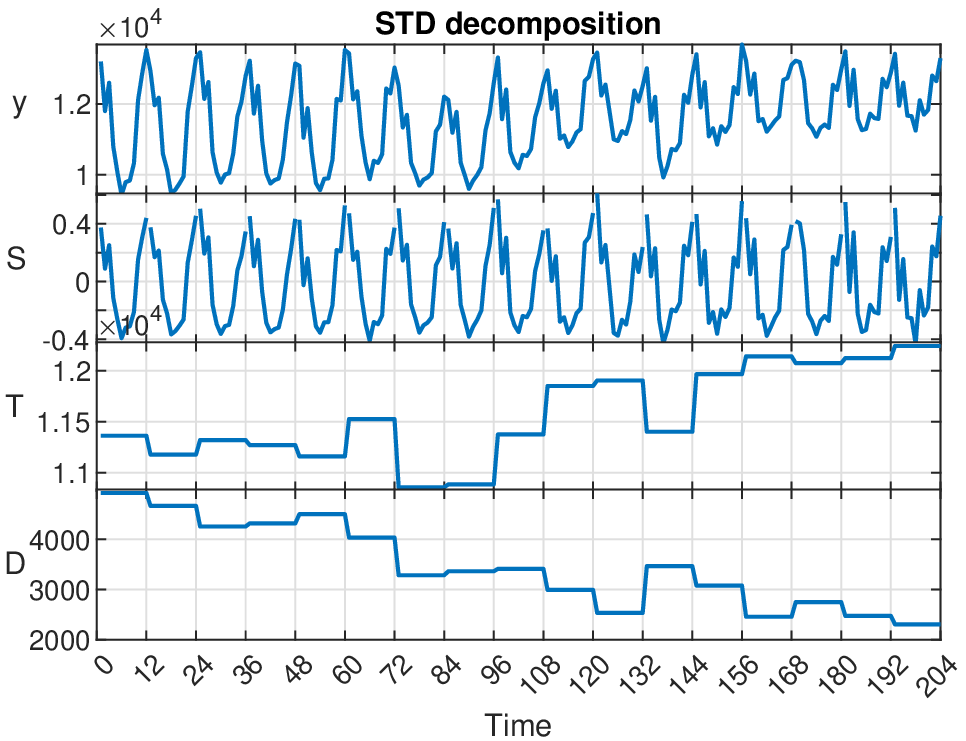}
	\includegraphics[width=0.16\textwidth]{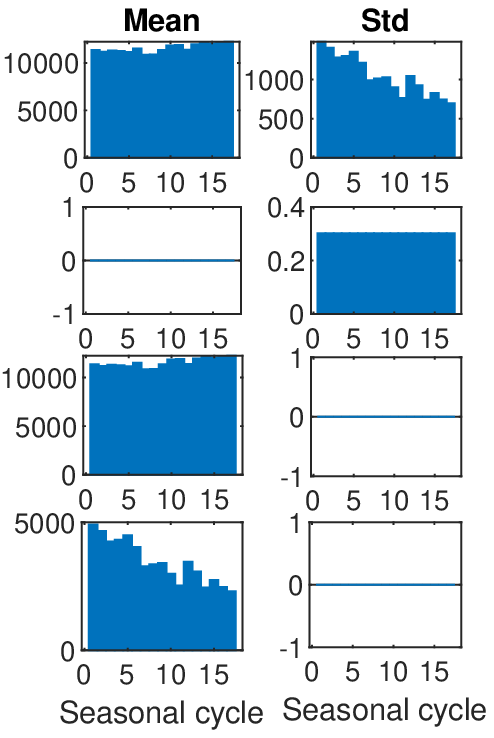}
	\caption{Monthly electricity demand time series decomposition using STD.}
	\label{STD}
\end{figure}

Fig. \ref{STD} shows an example of STD decomposition of the time series of monthly electricity demand for Poland. Note that the trend and dispersion components are step functions, where the step length corresponds to seasonal period $n$. The trend expresses the level of the time series in successive seasonal periods, while the dispersion expresses the variation of the time series elements in these periods. The seasonal component is composed of the seasonal patterns, which are centered, i.e. their average value is zero, and unified in variance, i.e. their dispersion is the same. Moreover, when we express seasonal patterns by vectors, $\mathbf{s}_i=[S_{i,1}, ..., S_{i,n}]$, where $S_{i,j}$ is the $j$-th component of the $i$-th seasonal pattern, their length is equal to one. Thus, they are normalized vectors. Although unified, the seasonal patterns differ in "shape". Their "shapes" express unified variations of the series in the successive seasonal periods. Note that the "shapes" are not smoothed or averaged as in the standard decomposition methods.

A variant of STD is STD with a reminder component, STDR, defined as follows:

\begin{equation}
y_t = S'_t \times D_t+T_t+R_t
\label{eqd2}
\end{equation}
where $S'_t$ is an averaged seasonal component and $R_t$ is a reminder component.

In STDR, the trend and dispersion components are defined in the same way as in STD. The seasonal component is defined using an average seasonal pattern, $\{\bar{S}_j\}_{j=1}^n$, determined as follows:

\begin{equation}
\bar{S}_j=\frac{1}{K}\sum_{i=1}^K S_{i,j}
\label{ss}
\end{equation}

The seasonal component in STDR is a sequence of $K$ averaged seasonal patterns:  

\begin{equation}
\{S'_t\}_{t=1}^N=
\{\underbrace{\{\bar{S}_j\}_{j=1}^n, ..., \{\bar{S}_j\}_{j=1}^n}_\text{$K$ times}\}
\label{ss1}
\end{equation}
thus, it is identical across all seasonal periods.

The reminder component is calculated from \eqref{eqd2}:
\begin{equation}
R_t = y_t - S'_t \times D_t+T_t
\label{rr}
\end{equation}

\begin{figure}[h]
	\centering
	\includegraphics[width=0.3175\textwidth]{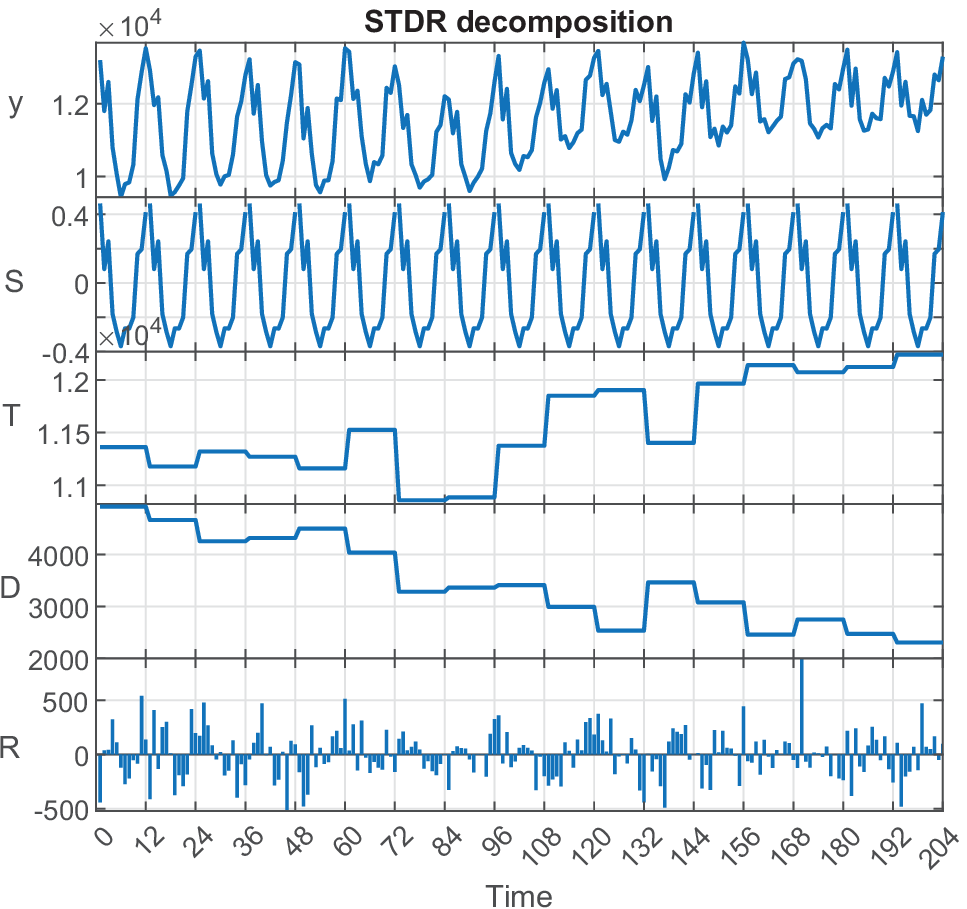}
	\includegraphics[width=0.1635\textwidth]{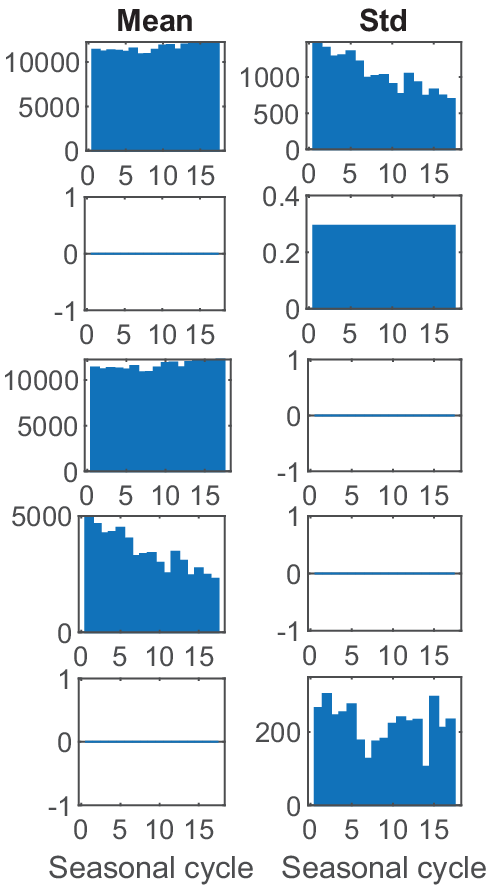}
	\caption{Monthly electricity demand time series decomposition using STDR.}
	\label{STDR}	
\end{figure}

\begin{figure}[h]
	\centering
	\includegraphics[width=0.35\textwidth]{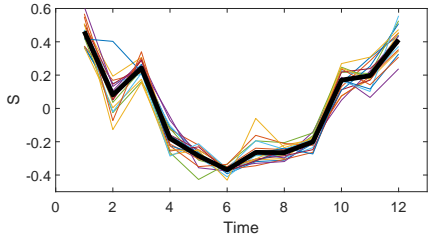}
	\caption{Seasonal patterns of the monthly electricity demand time series (averaged pattern drawn with a thick line).}
	\label{seas}	
\end{figure}

\begin{figure}[h]
	\centering
	\includegraphics[width=0.45\textwidth]{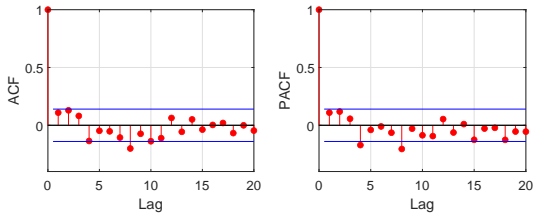}
	\caption{ACF and PACF plots for the reminder component of STDR applied for the monthly electricity demand time series.}
	\label{ACF}	
\end{figure}

An example of STDR decomposition is depicted in Fig. \ref{STDR}. Note the same trend and dispersion components as in Fig. \ref{STD} for STD, and the different seasonal component, which for STDR is composed of the same averaged seasonal pattern. Fig. \ref{seas} shows the seasonal patterns and the averaged pattern. The remainder correspond to the mismatch between the original seasonal cycles and the averaged seasonal cycles. Thus, it contains additional dispersion resulting from averaging the seasonal cycles. This dispersion is lower for the cycles whose patterns are similar to the averaged pattern. Note that the reminder has a zero average value in each seasonal period. To assess its stationarity visually, Fig. \ref{ACF} shows the plots of its sample autocorrelation function (ACF) and sample partial autocorrelation function (PACF). As can be seen from this figure, 
most of the spikes are not statistically significant, i.e. the reminder series is not highly correlated, which characterizes a stationary process. To confirm that the reminder is stationary, we apply three formal tests for a unit root in a univariate time series: augmented Dickey-Fuller test, Kwiatkowski, Phillips, Schmidt, and Shin test, and Phillips-Perron test. All tests confirmed stationarity at a 1\% level of significance.

Remark: The dispersion component can be defined using a standard deviation as a diversity measure (which is diversity \eqref{dy} divided by $\sqrt{n}$). In such a case, all components including the remainder have the same shape as in the standard formulation, but the dispersion component decreases its range $\sqrt{n}$ times, and the seasonal component increases its range $\sqrt{n}$ times. 

\section{Application Examples}

In this section, we apply the proposed decomposition method to time series of different nature including multiple seasonality and no seasonality. We also present forecasting approaches based on STD decomposition.

\subsection{Time Series Analysis using STD}

\begin{figure}[ht]
	\centering
	\includegraphics[width=0.38\textwidth]{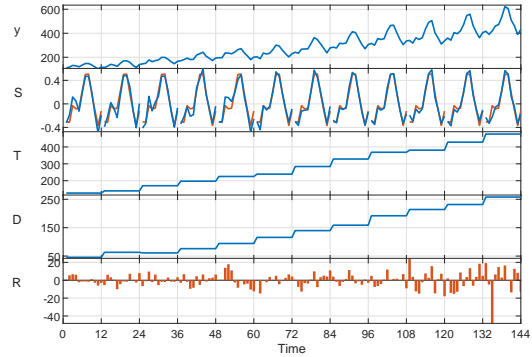}
	\caption{Airline passengers time series (in thousands) and its decomposition using STD and STDR.}
	\label{STDar}	
\end{figure}

As a first example we use the classic Box \& Jenkins airline data \cite{Box16}, i.e. monthly totals of international airline passengers from 1949 to 1960. This time series expresses an increasing trend and strong yearly periodicity ($n=12$) that increases in size with the series level -- see top panel in Fig. \ref{STDar}. Fig. \ref{STDar} shows both STD and STDR decompositions. They have the same trend and dispersion components. The seasonal component for STD is shown in blue, and the seasonal component for STDR as well as the reminder component are shown in red. Note that the seasonal patterns generated by STD are very similar in shape. 

Table \ref{tab1} shows the results of stationarity tests for the reminder, i.e. augmented Dickey-Fuller test (aDF), Kwiatkowski, Phillips, Schmidt, and Shin test (KPSS), and Phillips-Perron test (PP). All the tests confirm stationarity with 1\% significance level. Table \ref{tab1} also shows the median and interquartile range of the ratio of the reminder to the time series defined as follows:

\begin{equation}
r_t = \left|\frac{R_t}{y_t} \right|*100
\label{mra}
\end{equation}
The ratio of the reminder to the time series for Airline data is relatively small, 1.78\%.

\begin{table}[ht]
	\caption{Results of the stationarity tests for the reminder and the ratio of the reminder to the time series.}
	\begin{tabular}{lcccc}
		\toprule
		Data  & aDF & KPSS & PP & Median$(r_t) \pm$ IQR$(r_t)$  \\
		\midrule    
Airline                        & +   & +    & +  & $1.78 \pm 2.26$                   \\
Unemployment                   & +   & +    & +  & $2.29 \pm 3.24$                   \\
S\&P 500         & +   & +    & +  & $1.12 \pm 1.46$                   \\
S\&P 500 returns & +   & +    & +  & $92.96 \pm 72.57$                 \\
Electricity                    & +   & +    & +  & $2.04 \pm 2.95$                   \\
Mackey-Glass                   & +   & +    & +  & $\phantom{0}8.87 \pm 12.31$                 \\

		\bottomrule
	\end{tabular}
	\label{tab1}

\end{table}

\begin{figure}[ht]
	\centering
	\includegraphics[width=0.38\textwidth]{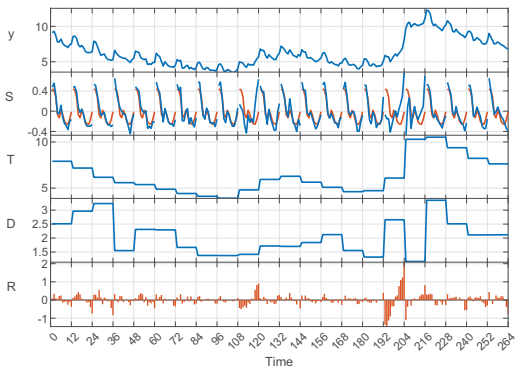}
	\caption{Unemployment time series decomposition using STD and STDR.}
	\label{STDun}	
\end{figure}

The second example uses data for the US unemployment rate for males (16 years and over) observed from January 1992 to December 2013 ($n=12$). This series was analysed extensively in \cite{Dag16}. It exhibits yearly seasonality with strong asymmetric behavior, i.e. it displays steep increases that end in sharp peaks and alternate with much more gradual and longer declines that end in mild troughs \cite{Gro17}. Thus the seasonal patterns are generally similar to each other. The seasonal patterns observed in Fig. \ref{STDun} are similar in shape, except for three patterns, which reflect sharp spikes in unemployment in the final months of the year, i.e. sequences 109-120, 193-204 and 205-2016. Due to a deviation from the typical shape for these three sequences, the reminder takes larger values for them than for other annual sequences. Nevertheless, it passes the stationarity tests, see Table \ref{tab1}. The ratio of the reminder to the time series for unemployment data is 2.29\%. 


\begin{figure}[ht]
	\centering
	\includegraphics[width=0.38\textwidth]{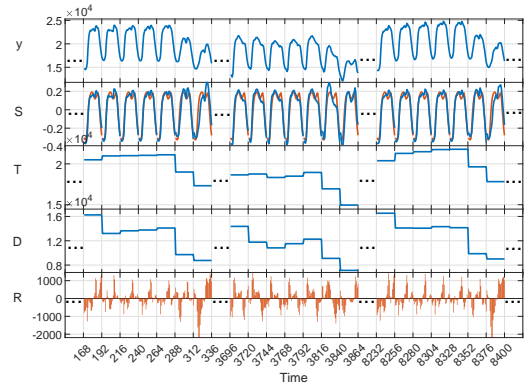}
	\caption{Hourly electricity demand time series decomposition using STD and STDR.}
	\label{STDstlf}	
\end{figure}

The third example concerns hourly electricity demand. Time series of this type exhibit triple seasonality: yearly, weekly and daily. 
The seasonalities are related to the local climate, weather variability and the nature of a country's economy. Fig. \ref{STDstlf} shows decomposition products of the hourly electricity demand for Poland in 2018. We assumed a seasonal period as a daily one ($n=24$). In Fig. \ref{STDstlf}, we show three weekly sequences of the time series, from January, July and December. As can be seen from this figure, the seasonal component expresses daily patterns whose shapes are related to the day of the week and season of the year. The daily patterns representing the working days from Tuesday to Friday for the same period of the year are similar to each other. Patterns for Mondays are slightly different from them. Saturdays and Sundays have their own shapes. Note that the trend and dispersion components have both weekly and yearly seasonalities. These two components can be further decomposed using STD or STDR. The ratio of the reminder to the time series is only 2.04\%. The reminder passes all the  stationarity tests.  
  
\begin{figure}[ht]
	\centering
	\includegraphics[width=0.38\textwidth]{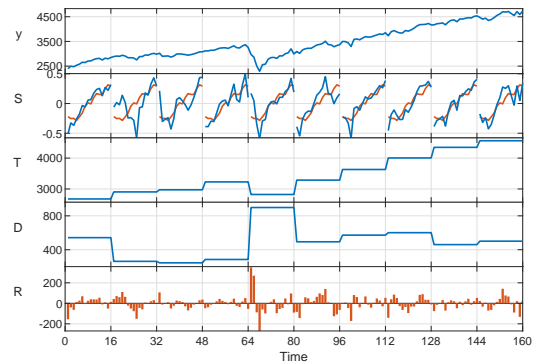}
	\caption{S\&P 500 Index time series decomposition using STD and STDR.}
	\label{STDsp}
\end{figure}

The next two examples are for financial time series. We analyse one of the most important stock market indexes, S\&P 500. It tracks the performance of the 500 largest companies listed on stock exchanges in the United States. 
Fig. \ref{STDsp} shows decomposition of the weekly S\&P 500 Index over the period 2019-2021. S\&P 500 Index shows development within a rising trend that dips at the beginning of 2020 due to the Covid-19 crisis. The time series does not express seasonality. We assume $n=16$ weeks for STD decomposition. Because of the rising trend, the 16-week patterns forming the seasonal component have a rising character, but differ due to significant random noise. For the pattern representing the Covid-19 fall period (sequence 65-80) the highest remainder values are observed as well as the highest dispersion. The ratio of the reminder to the time series is low, 1.12\%. The reminder passes all stationarity tests (see Table \ref{tab1}).

\begin{figure}[ht]
	\centering
	\includegraphics[width=0.38\textwidth]{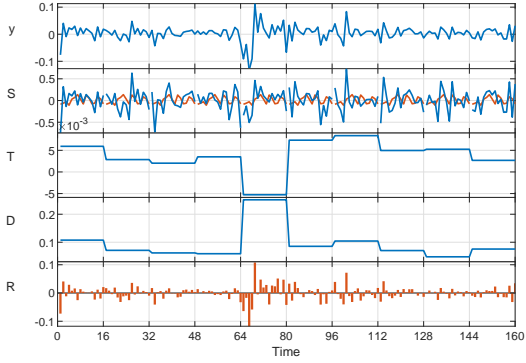}
	\caption{Decomposition of S\&P 500 weekly log returns using STD and STDR.}
	\label{STDrsp}
\end{figure}

Fig. \ref{STDrsp} shows decomposition of the S\&P 500 returns calculated as $\ln\left(y_t/y_{t-1}\right)$, where $y_t$ represents the original time series. While the original time series of S\&P 500 Index is nonstationary, the returns fluctuate around a stable mean level \cite{Box16}. However, their variability around the mean changes. In the period 2019-21, it is highest during the Covid-19 crisis, see Fig. \ref{STDrsp}, where the dispersion and remainder are highest for the crisis period, i.e. sequence 65-80. The ratio of the reminder to the time series is high (around 93\%), which indicate the dominant content of the noise in the series of returns. The reminder passes all the stationarity tests (see Table \ref{tab1}). 

The last example concerns decomposition of a synthetic time series -- a Mackey-Glass time series which is produced by the delay differential equation \cite{Mac77}, i.e. $\frac{dx(t)}{dt}=\frac{ax(t-\tau)}{1+x^{10}(t-\tau)}-bx(t)$, where we assumed typical values for parameters: $a=0.2$, $b=0.1$, $x(0)=1.2$, and $\tau=17$. With these parameters, the time series is chaotic and exhibits a cyclic behavior. This time series is commonly used as a benchmark for testing different forecasting methods, because it has a simple definition, and yet its elements are hard to predict \cite{Pal05}. 

\begin{figure}[ht]
	\centering
	\includegraphics[width=0.38\textwidth]{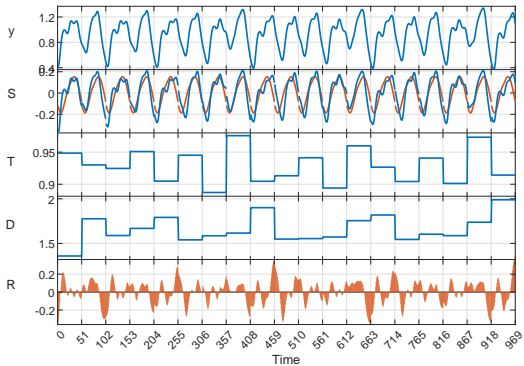}
	\caption{Mackey-Glass time series decomposition using STD and STDR.}
	\label{STDmg}
\end{figure}

Fig. \ref{STDmg} depicts the Mackey-Glass time series decomposition. The series was computed with a time sampling of 1. The sequence for $t$ ranging from 101 to 1070 is shown. We assumed a seasonal pattern length as $n=51$. Note the irregular character of the seasonal patterns and also the chaotic variability in the trend and dispersion components. The ratio of the reminder to the time series is 8.87\%. The reminder passes all the stationarity tests (see Table \ref{tab1}).    

\subsection{Time Series Forecasting using STD}

Decomposition helps to improve understanding of the time series, but it can also be used to improve forecast accuracy. Extracted components have lower complexity than the original time series and so can be modelled independently using simple models. In the case of STDR, the seasonal pattern does not change and we can use a naive approach to predict it for the next period. In STD, the seasonal pattern changes and we should use an appropriate forecasting method to predict it. Note that in the examples considered above the reminder was a stationary process. Thus it can be predicted even by those methods that require stationarity such as ARIMA. 
Trend and dispersion components can exhibit seasonality such as in the case of hourly electricity demand shown in Fig. \ref{STDstlf}. Such time series can be predicted using seasonal models or can be further decomposed into simple components using STD or STDR.   

To avoid the troublesome task of forecasting all the components extracted by STD, in \cite{Dud15a}, a method was described which combines all components into an output pattern (in fact in \cite{Dud15a} many input and output patterns were proposed. We focus on the patterns denoted as X3.1 and Y3.1, which are related to STD). The forecasting model predicts output patterns based on the input patterns which are seasonal patterns expressed by vectors $\mathbf{s}_i=[S_{i,1}, ..., S_{i,n}]$, where $S_{i,j}$ is the $j$-th component of the $i$-th seasonal pattern. They are defined as follows (this is an alternative notation to \eqref{Se}):     

\begin{equation}
\mathbf{s}_i=\frac{\mathbf{y}_{i}-\bar{y}_i}{\tilde{y}_i}
\label{Si}
\end{equation}
where $\mathbf{y}_i=[y_{i,1}, ..., y_{i,n}]$ is a vector representing the $i$-th seasonal sequence of the time series. 

Thus, the input patterns are centered and normalized seasonal sequences. The output pattern represents a forecasted seasonal pattern. It is defined as:

\begin{equation}
\mathbf{s}_{i+\tau}=\frac{\mathbf{y}_{i+\tau}-\bar{y}_i}{\tilde{y}_i}
\label{Sit}
\end{equation}
where $\mathbf{s}_{i+\tau}=[S_{i+\tau,1}, ..., S_{i+\tau,n}]$ and $\tau \geq 1$ is a forecast horizon.

\begin{figure}[ht]
	\centering
	\includegraphics[width=0.45\textwidth]{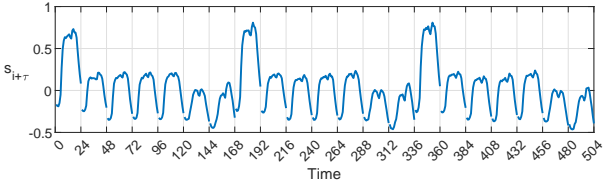}
	\caption{Examples of output patterns for hourly electricity demand time series (first day is Monday, last day is Sunday).}
	\label{out}
\end{figure}

Note that in \eqref{Sit} to calculate the output pattern, we use the average and dispersion for sequence $i$ and not for sequence $i+\tau$. This is because these two coding variables for future sequence $i+\tau$, which has been just forecasted, are not known. Using the coding variables for the previous period has consequences: the output patterns are no longer centered and normalized vectors like the input patterns are. But if the mean value of the series and its dispersion do not change significantly in the short period, i.e. $\bar{y}_{i+\tau} \approx \bar{y}_i$ and $\tilde{y}_{i+\tau} \approx \tilde{y}_i$, the output patterns are close to centered and normalized. For time series with multiple seasonality, we cannot assume that the trend and dispersion are constant in the short term because they are influenced by additional seasonal fluctuations. For example, the average values and dispersions of daily sequences can changes with the weekly seasonality, see Fig. \ref{STDstlf}. This translates into output patterns. Referring to the example shown in Fig. \ref{STDstlf}, the output patterns for Mondays are coded with the averages and dispersions of Sunday sequences (for $\tau=1$), which are lower than those for Mondays. This has the effect of shifting the output patterns for Monday up and stretching it. For similar reasons, output patterns for Saturdays and Sundays are placed lower than output patterns for the other days of the week and are less stretched (compare this in Fig. \ref{out}). Thus, the output patterns are not unified globally but are unified in groups composed of the same days of the week (unified means that they have a similar average value and dispersion). For this reason, it is reasonable to construct the forecasting models that learn from data representing the same days of the week. For example, when we train the model to forecast the daily sequence for Monday, a training set for it is composed of the output patterns representing all Mondays from history and the corresponding input patterns representing the previous days (depending on the forecast horizon; Sundays for $\tau=1$).

The forecasting model fits function $f: \mathbf{s}_i \to \mathbf{s}_{i+\tau}$. The forecasted output pattern, $\hat{\mathbf{s}}_{i+\tau}$, is postprocessed to obtain the real forecasted sequence using transformed equation \eqref{Sit}:

\begin{equation}
\hat{\mathbf{y}}_{i+\tau}=\hat{\mathbf{s}}_{i+\tau}\tilde{y}_i+\bar{y}_i
\label{yit}
\end{equation}
Note that in \eqref{yit}, the coding variables, $\bar{y}_i$ and $\tilde{y}_i$, are known from the most recent history. This enables us to perform the postprocessing (decoding). 

\begin{figure}[ht]
	\centering
	\includegraphics[width=0.48\textwidth]{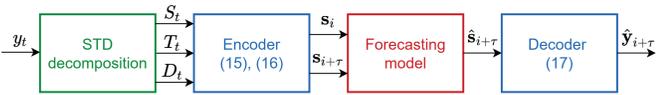}
	\caption{Diagram of forecasting based on STD.}
    \label{For}
\end{figure}

Note that equations \eqref{Si} and \eqref{Sit} filter out the current process variability from the data, i.e. filter out the local average and dispersion. The model learns on filtered (unified) patterns and forecasts the output pattern $\hat{\mathbf{s}}_{i+\tau}$. Equation \eqref{yit} introduces information about the process variability in sequence $i$ (the most recent historical sequence) into the output data. This approach, depicted in Fig. \ref{For}, enables us to take into account the local variability of the process when constructing the forecast. 

Due to representation of the time series by unified patterns $\mathbf{s}_i$ and $\mathbf{s}_{i+\tau}$, the forecasting problem simplifies and can be solved using simple models. The models proposed in \cite{Dud15b} and \cite{Dud21} are based on the similarity between the patterns. They assume that similarity in the input space is related to the similarity in the output space. Thus the forecasted output pattern is constructed from the training output patterns paired with the most similar input training patterns to the query pattern. To model function $f$, which in this approach has a nonparametric form, many models has been investigated such as the
nearest-neighbor model, fuzzy neighborhood model,
kernel regression model, general regression neural network, and  pattern clustering-based models (including classical clustering methods and artificial immune systems).

In \cite{Dud16}, function $f$ was modeled locally using different linear models including stepwise and lasso regressions, principal components regression and partial least-squares regression. In \cite{Dud15c}, a random forest was used to model $f$, and in \cite{Dud16b}, different neural network architectures were compared. In \cite{Dud20}, it was shown that STD decomposition improves forecasting accuracy of an advanced hybrid and hierarchical deep learning model which combines exponential smoothing and residual dilated long short-term memory network.

\subsection{Discussion}

The advantage of STD over the standard decomposition methods is that it extracts a dispersion component showing short-term variability of the time series over time, i.e. variability of the series in seasonal periods. This is very useful for analysing heteroscedastic time series, which are very common in different domains such as finance, business, industry, meteorology etc. The dispersion component enables direct evaluation of the series variance, which is not allowed by standard methods, where this variance is expressed in many components, and additional calculations are required to estimate it.

The seasonal component extracted by STD is composed of normalized seasonal patterns. They represent real seasonal cycles which are detrended and unified in variance. The normalized patterns emphasize the "shapes" of the seasonal cycles and make them easier to compare. Comparing the shapes of seasonal cycles is impossible when using standard decomposition methods. This is because these methods either average the seasonal cycles, like the classical additive and multiplicative methods and STL, or express these cycles in many components, such as wavelet decomposition and EMD. Being able to compare the seasonal cycle shapes can be very useful for evaluating the stability of the seasonal pattern or its change over time. For example we can observe how the shape of the daily electricity demand curve differs for different days of the week and changes from season to season or year to year. We can also compare the shapes for different countries.    

STDR averages the normalized seasonal patterns and delivers the reminder component. This component expresses the difference between real time series and the series with unified seasonal cycles, i.e the series which has the same trend and dispersion components as the real series but its seasonal pattern is averaged. Analysing the reminder, we can detect periods in which the seasonal patterns differ from the averaged pattern most. For example, the unemployment time series shows
increased differences in the shapes of seasonal cycles in periods 109-120 and 193-204. In these periods, the falling series temporarily increases (see Fig. \ref{STDun}). Patterns in the reminder can be further investigated in order to analyze the magnitudes and directions of deviations of seasonal cycles from the averaged cycles.

It is worth emphasizing the high interpretability of STD. It extracts easy to understand and informative components expressing the main properties of the series, i.e. tendency of the series (trend averaged in seasonal periods), local variability of a series (dispersion in seasonal periods) and shapes of a seasonal cycles (unified seasonal patterns). Compared to STD components, the components produced by standard methods, such as high frequency IMFs and details, are not easy to interpret. They do not express clear patterns.  

Another very important advantage of STD and STDR are their simple algorithms, which can be coded in less then 30 lines of code in Matlab, as shown in Appendix A. The algorithms do not require complex computation. The components can be extracted using simple formulas (see mathematical formulation composed of just three equations for STD: \eqref{my}, \eqref{dy} and \eqref{Se}, and additional two for STDR: \eqref{ss}, \eqref{rr}). 
Note that both versions, STD and STDR, have no parameters when used for seasonal time series. For non-seasonal series only one parameter should be selected, i.e. the "seasonality" period $n$. The simplest methods among the standard methods, the classical additive and multiplicative methods, require selection of one parameter, i.e. the order of the moving average. More sophisticated methods, such as STL, wavelet decomposition and EMD, require more parameters. For STL these include: the spans of the Loess windows for trend, seasonality and low-pass filter, and the degrees of the locally-fitted polynomials for trend, seasonality and low-pass filter. Wavelet decomposition requires the
number of decomposition levels and wavelet type (or alternatively the coefficients of the low-pass and high-pass filters), while EMD requires selection of the interpolation method for envelope construction, decomposition stop criteria and shifting stop criteria. EMD suffers from a boundary problem which results in anomalously high amplitudes of the IMFs and artifact wave peaks towards the boundaries \cite{Sta20}. Another boundary problem occurs for classical additive and multiplicative decompositions. Due to the need to estimate the moving average using the two-sided window, the estimate of the trend and reminder are unavailable for observations near boundaries. In the proposed STD and STDR there are no boundary problems.  

Although STD and STDR were designed for time series with single seasonality, they can be used for non-seasonal time series. In such a case the seasonal component does not express a regular pattern such as for S\&P 500 returns (see Fig. \ref{STDrsp}) or expresses a pattern resulting from the general tendency of the time series such as for S\&P 500 Index, where the rising "seasonal" patterns reflect the rising trend of the series (see Fig. \ref{STDsp}). STD and STDR can also be useful for decomposition of time series with multiple seasonality. In such a case, the seasonal component expresses the seasonal patterns of the shortest period, and trend and dispersion components express seasonalities of the longer periods, see example in Fig. \ref{STDstlf}. To extract all seasonal components, the STD/STDR decomposition can be applied for trend and dispersion components again.

Based on STD decomposition, we can define the input and output variables for the forecasting models. The input variables are just the seasonal patterns for period $i$, while the output variables are the seasonal cycles for period $i+\tau$ encoded using the average and dispersion for period $i$. Such encoding of both input and output variables filters out the trend and variability of the time series. This makes the relationship between the variables simpler. Thus this relationship can be modeled using simpler models such as linear regression or similarity-based models. Forecasting models using STD-based coding are great at dealing with time series with multiple seasonality, which has been proven in many papers \cite{Dud15a, Dud15b, Dud16, Dud15c, Dud16b}.

\section{Conclusion}

Time series decomposition into several components representing an underlying pattern category is a key procedure for time series analysis and forecasting. In this work, we propose a new decomposition method, seasonal-trend-dispersion decomposition. It has two variants: with (STDR) and without (STD) the reminder component. 
The proposed decomposition can be summarized as follows:

\begin{enumerate}
  \item It distinguishes itself from existing methods in that it extracts the dispersion component which expresses the short-term variability of the time series. A separate dispersion component is very useful for heteroscedastic time series analysis.    
  \item It produces interpretable components which express the main properties of the time series: the trend, dispersion and seasonal patterns.
  \item In STD, a seasonal component is composed of centered and normalized seasonal patterns, which express the "shapes" of the seasonal cycles. By emphasizing these shapes, STD facilitates comparison and analysis of the seasonal cycles. 
  \item In STDR, a remainder component expresses the difference between the real seasonal cycles and the averaged cycles. It enables the detection of outlier seasonal cycles that differ in shape from the averaged cycles.
  \item It has no parameters to adjust for seasonal time series. For non-seasonal time series, only one parameter should be selected.
  \item The algorithms of STD and STDR are very simple and easy to implement. The computation time is very fast.
  \item STD can be used for encoding the input and output variables for the forecasting models. STD-based encoding simplifies the relationship between variables which translates into simpler models and improved forecasting accuracy.
\end{enumerate}

\appendices
\section{STD Implementation.} \label{listening}

The source code is available here: \url{https://github.com/GMDudek/STD}.

\begin{lstlisting}
function [S,T,D,R] = STD_decomposition(data,period,is_reminder)
% STD_decomposition - seasonal-trend-dispersion decomposition of time series
%
% data - time series, 1 x N or N x 1
% period - length of the seasonal cycle
% is_reminder - 0 for STD (without reminder), 1 for STDR (with reminder)
% S - seasonal component, 1 x N
% T - trend component, 1 x N
% D - dispersion component, 1 x N
% R - reminder component, [] for STD, 1 x N for STDR

if iscolumn(data)
    y = data';
else
    y = data;
end
n = period;
N = length(y);
K = N/n;

if rem(N,n)
    disp(['Length of the series (', num2str(N),') should be a multiple of the seasonal period (', num2str(n),')']);
end

yy = reshape(y,n,K);

%trend
ym = mean(yy); 
q = repmat(ym,n,1); 
T = q(:)'; 

%dispersion
yd = std(yy)*n^0.5; 
q = repmat(yd,n,1); 
D = q(:)'; 

%seasonal for STD
S = (y - T)./D; 
S = S(:)'; 

%reminder for STD
R = []; 

%seasonal and reminder for STDR
if is_reminder 
    q = reshape(S,n,K);
    sp = mean(q,2)';
    S = repmat(sp,1,K);
    R = y - (S.*D + T);
end
\end{lstlisting}

\begin{IEEEbiography}[{\includegraphics[width=1in,height=1.25in,clip,keepaspectratio]{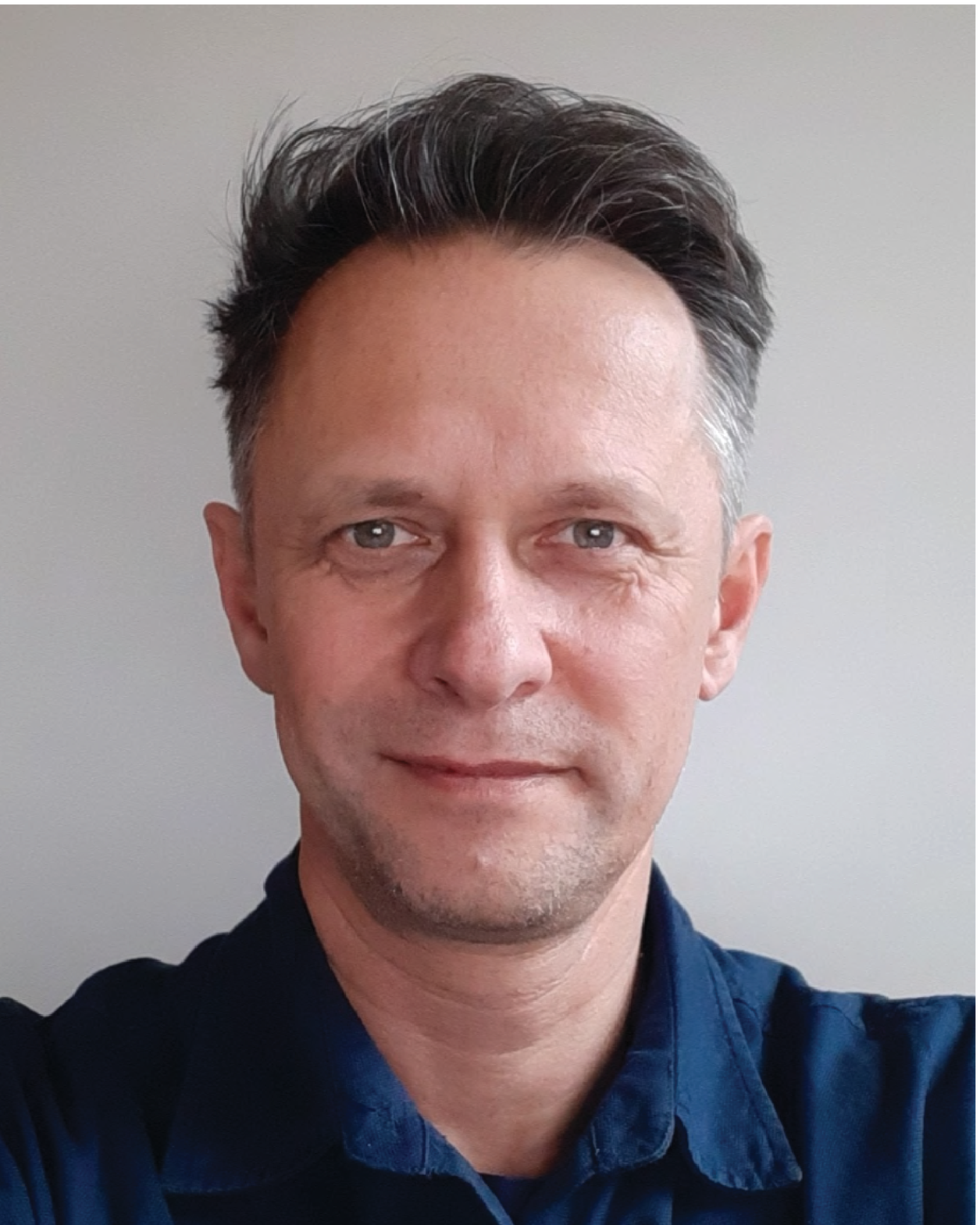}}]{Grzegorz Dudek}
	received his PhD in electrical engineering from Czestochowa University of Technology (CUT),
	Poland, in 2003 and habilitation in
	computer science from Lodz University of
	Technology, Poland, in 2013. Currently,
	he is an associate professor at the
	Department of Electrical Engineering,
	CUT. He is the author of
	two books concerning machine learning methods for load
	forecasting and evolutionary algorithms for unit commitment
	and over 120 scientific papers. He came third in the Global Energy Forecasting Competition 2014 (price forecasting track). His research interests include
	pattern recognition, machine learning, artificial intelligence,
	and their application to practical classification, regression, forecasting
	and optimization problems. 	
\end{IEEEbiography}




\end{document}